\documentclass[showpacs,aps,pre,superscriptaddress,floatfix]{revtex4}
\usepackage{graphicx} 

\begin{document}

\title{Non-universal disordered Glauber dynamics}

\author{Marcelo D. Grynberg} 

\affiliation{Departamento de F\'{\i}sica, Universidad Nacional de  La Plata, 
1900 La Plata, Argentina}

\author{Robin B. Stinchcombe}

\affiliation{Rudolf Peierls Centre for Theoretical  Physics, University of Oxford,
1 Keble Road, Oxford OX1 3NP, UK}

\begin{abstract}
We consider the one-dimensional Glauber dynamics with coupling disorder in 
terms of bilinear fermion Hamiltonians. Dynamic exponents embodied in the 
spectrum gap of these latter are evaluated numerically by averaging over both 
binary and Gaussian disorder realizations. In the first case, these exponents are 
found to follow the non-universal values of those of plain dimerized chains. In the 
second situation their values are still non-universal and sub-diffusive below a 
critical variance above which, however, the relaxation time is suggested to grow 
as a stretched exponential of the equilibrium correlation length.
\end{abstract}


\maketitle

\section{Introduction}

In the absence of general principles accounting for the evolution of non-equilibrium
systems, kinetic Ising models have been used to characterize a large variety of slow
non-equilibrium processes \cite{Marro, Privman}. Among the  most ubiquitous of
these latter, the coarsening dynamics after a sudden quench from a disordered phase
at high temperatures to an ordered one below the critical point still remains a subject
of much interest and debate \cite{Bray}. For {\it pure} substrates it is by now
established that coarsening domains are  characterized by a single time-dependent
length scale growing as $t^{1/z}$, where $z$ is the dynamic exponent of the 
universality class to which the dynamics belongs. Depending on whether or not the 
order parameter is preserved by this latter, e.g. Kawasaki \cite{Kawa} or Glauber 
\cite{Glauber} dynamics, the network of (ferromagnetic) domains will coarsen
respectively according to a sub-diffusive Lifshitz-Slyozov growth ($z = 3$, see however
Ref.\cite{Grynberg}\,), or following a plain diffusive Allen-Cahn behavior ($z = 2$)\, 
\cite{Bray}.

On the other hand, the case of coarsening systems with quenched disorder substrates
modelling the effects of various experimental situations has also received considerable
attention \cite{Bray, Huse}, especially in the context of random field and random
bond Ising models \cite{Forgacs, Biswal, Paul, Corberi, Lippiello}. In general, it is
expected that quenched impurities play the role of energy barriers slowing down
coarsening domains, the boundaries of which move by thermal activation over the
landscape of these trapping centers \cite{Huse}. However, it is not yet clear whether
universal growth can take place at late evolution stages in the presence of strong
disorder.  Recent Monte Carlo simulations \cite{Lippiello} of the Glauber dynamics
using random ferromagnetic couplings drawn from uniform probability distributions
suggest that after long pre-asymptotic crossovers the final growth in one dimension
(1D) recovers the usual diffusive behavior referred to above, while becoming
logarithmically slow in the two-dimensional case. For the 1D situation this is somehow
intriguing as already at the level of a plain alternating-bond or dimerized chain
\cite{Droz} the dynamic exponents are known to be non-universal and sub-diffusive 
($ z > 2$).

As part of the ongoing efforts in this context, here we revisit the 1D Glauber disordered
dynamics in terms of bilinear fermion fields associated to the kinetics of its kinks or
domain walls. Here, we rather focus on the dynamic exponent characterizing the 
actual growth of relaxation times $\tau$ with equilibrium correlation lengths $\xi$, 
which according to critical dynamic theories \cite{Hohenberg} should scale as $\tau 
\propto \xi^x$ . But it is known \cite{Privman, Menyhard} that this latter exponent 
coincides with $z$ because at times comparable with $\tau$ the average domain size
becomes of order $\xi$, so both descriptions are ultimately equivalent at large scales
of space and time. The idea is to avoid the problem of dealing with prohibitively 
long transient regimes by pinpointing directly the relaxation time embodied in the 
spectrum gap of the evolution operator which, for that purpose, will be constructed 
and diagonalized in a domain wall representation. Although the critical point is 
strictly zero, the analysis is kept within low temperature regimes (with $\xi$ and 
$\tau$ becoming both arbitrarily large there), otherwise the dynamics would be 
rapidly arrested \cite{Derrida} owing to the proliferation of large quantities of 
metastable  states. The disorder considered throughout may include either exchange 
couplings of a single type (ferro or antiferromagnetic) or mixed ones, though as we 
shall see in a moment, the dynamics of these situations can be mapped onto each
other via a simple spin transformation.

The outline of this work is organized as follows. In Sec. II we describe the basic kinetic
steps under random spin exchanges using a kink or dual representation, and write 
down the underlying Glauber generator in quantum spin notation. Exploiting detailed 
balance \cite{Kampen} we then recast this latter operator in terms of bilinear and
symmetric fermion forms, thus allowing to evaluate its spectra through the solution 
of a secular problem whose dimensions grow linearly with the system size. In Sec. III 
we present the results arising from the numerical diagonalizations of this problem 
both for binary and Gaussian disorder realizations. Non-universal dynamic exponents
as well as scaling forms drawn from the exact solution of the dimerized chain are 
then proposed and seen to be consistent with some of our numerical findings.
However, when the width of the Gaussian disorder exceeds a certain threshold the
relaxation time grows in a rather different (non-power-law) and much faster form. 
We close with Sec. IV which contains a recapitulation along with brief remarks on 
open issues and possible extensions of this work.

\section{Dynamics and bilinear forms}

Let us consider the Glauber dynamics of the Ising chain Hamiltonian ${\cal H} = 
- \sum_i \,J_i  \,S_i\, S_{i+1}\,$ with $L$ spins $S_i = \pm 1\,$ and disordered 
couplings $J_i$ chosen arbitrarily with any sign and strength. Unless stated otherwise,
periodic boundary conditions (PBC) will be used throughout. The system is in contact 
with a heat bath at temperature $T$, which causes the states  $\vert S\, \rangle  = 
\vert S_1,\,\dots\,, S_L \rangle$ to change by random flipping of single spins.  The  
probability per unit time $W ( S_i \to  - S_i)$ or transition rate for the $i$-th spin to 
flip is set to satisfy detailed balance \cite{Kampen} (see below), so that the system 
relaxes to equilibrium. As is known, the simplest choice of rates complying with this 
condition corresponds to that of Glauber \cite{Glauber}
\begin{eqnarray}
\nonumber
W ( S_i \to  - S_i)  &=& \alpha \left[\,1\, - \, \frac{S_i}{2}\, \left (\,
\gamma_i^- \,S_{i-1}\, + \,\gamma_i^+ \,S_{i+1}\, \right)\,\right]\,, \\
\label{g-rates}
\gamma_i^{\pm} & \equiv & \tanh \left( K_i + K_{i-1} \right) \pm 
\tanh \left( K_i - K_{i-1} \right)\,,
\end{eqnarray}
where $K_i \equiv J_i/k_BT$, and the spin flip rate $\alpha$ is hereafter taken as 
$1/2$. We shall always measure the temperature in energy units; equivalently the 
Boltzmann constant $k_B$ is set to 1. Note  that the nature of the dynamics is 
basically unaltered  by whether some or all couplings are ferromagnetic ($J_i > 0$) or
antiferromagnetic ($J_i < 0$). To check out rapidly this issue consider for simplicity a 
chain with open boundary conditions (just to avoid frustration due to eventual 
mismatch of antiferro exchanges), along with the mapping
\begin{equation}
\label{mapping}
R_1  \equiv  S_1\,,\;\; R_i  \equiv  S_i \,\prod_{n=1}^{i-1} {\rm sign} 
\left(J_n \right)\;\; {\rm for}\;\; 1 < i \le L\,,
\end{equation}
to new spins $R_i = \pm 1$. 
Then, to preserve the energy of  the mapped configurations and therefore to maintain
invariant all transition rates (these  in turn depending on $\Delta{\cal H}/T$), clearly
in the equivalent $R$-system all couplings must become ferromagnetic because 
$S_i \,S_{i+1} \to \,{\rm sign} \,( J_i) \, R_i \, R_{i+1}$. In particular, the dynamics of
the random $\pm J$ chain would then reduce to the homogeneous one. Also, notice
that the {\it single}-spin flip dynamics maps onto itself. In that latter respect, this 
argumentation would {\it not} apply to the Kawasaki dynamics \cite{Kawa} as owing
to the exchange of $S$-pairs under antiferro couplings, the mapping (\ref{mapping})
would then allow {\it parallel} $R$-pairs to flip.

Strictly, Ising Hamiltonians possess no intrinsic dynamics since all involved spin
operators commute with one another. But when these systems are endowed with 
extrinsic transition rates $W ( S \to S')$, such as those of Eq.\,(\ref{g-rates}), their 
time evolution is described in terms of a Markovian process governed by a master 
equation \cite{Kampen}. For our subsequent discussion it is convenient to think of
this latter as a Schr\"odinger equation in an imaginary time 
\begin{equation}
\partial_t \vert P (t)\, \rangle = - H\, \vert P (t) \,\rangle\,,
\end{equation}
under a pseudo Hamiltonian or evolution operator $H$. This provides the probability 
distribution to find the system in a state $\vert P(t) \,\rangle \equiv \sum_S P (S,t)\,
\vert S\, \rangle\,$ at time $t$, from the action of $H$ on a given initial distribution, 
i.e. $\vert P(t) \,\rangle = e^{- H\,t}\, \vert P(0)\,\rangle$. To ensure equilibrium at 
large times, the detailed balance condition referred to above simply requires that
$P_{eq} (S) \,W(S \to S') = P_{eq} (S') \, W(S' \to S), \;\forall \; \vert S \rangle,\vert S'
\rangle$. As usual, the diagonal and non-diagonal matrix elements of this Markovian
operator are given respectively by
\begin{equation}
\label{elements}
\langle \,S \vert H_d \vert S \,\rangle = \sum_{S'\ne S} W(S \to S')\,,\;\;
\langle\, S' \vert H_{nd} \vert S \,\rangle = -\,W(S \to S')\,.
\end{equation}
In particular, the first non-zero eigenvalue $E_1$ of such stochastic matrix  singles out
the relaxation time $\tau$ in which we are interested, i.e. $\tau =  1/{\rm Re}\; E_1 
> 0$, which is the {\it largest} characteristic time for any observable at a given low
temperature. The case of $E_0 = 0$ just corresponds to the stationary mode.

\begin{figure}[htbp]
\centering
\vskip -4.1cm
 \includegraphics[width=0.65\textwidth]{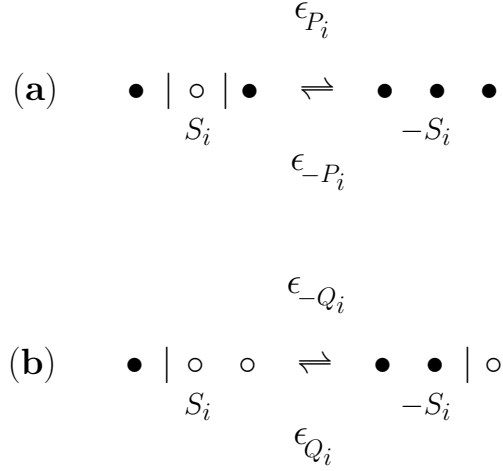}
\vskip -5.75cm
\caption{Disordered transition rates [\,Eq\,(\ref{rates})\,] for (a) creation-annihilation
of kink pairs , and (b) hopping of kinks. These latter are denoted by vertical lines 
separating domains of opposite spin orientations.}
\label{kinks}
\end{figure}

In order to build up and diagonalize the operational analog of Eq.\,(\ref{elements}),
in what follows it is convenient to work instead with the associated kink or domain 
wall dynamics rather than with the spin flipping process itself. Both representations 
are depicted schematically in Fig.\,\ref{kinks}.
Clearly from Eq.\,(\ref{g-rates}), the kink pairing and hopping rates indicated in this
diagram become respectively
\begin{eqnarray}
\nonumber
\epsilon_{_{\pm P_i}} &=&  \left (\, 1 \pm \tanh P_i \right)/2
\;\;\;,\,\;\; P_i = K_i + K_{i-1}\,,\\
\label{rates}
\epsilon_{_{\pm Q_i}} &=&  \left (\, 1 \pm \tanh Q_i \right)/2 
\;\;\;,\,\;\; Q_i = K_i - K_{i-1}\,.
\end{eqnarray}
Thus, if we think of the corresponding kink occupation numbers (0,1) as being related 
to usual spin-$\frac{1}{2}$ raising and lowering operators $\sigma^+,\sigma^-$, 
we can directly identify the non-diagonal operators  $H_{nd} \equiv H_{pairs} + 
H_{hops}$ associated to Eq.\,(\ref{elements}). Evidently, the transitions of these 
dual processes will be provided by
\begin{eqnarray}
\label{pairs}
H_{pairs} &=& - \sum_i\, \left (\, \epsilon_{_{-P_i}} \, \sigma^+_{i-1} \,
\sigma^+_i \,+\, \epsilon_{_{P_i}} \, \sigma^-_i \,   \sigma^-_{i-1} \,\right)\,,\\
\label{hops}
H_{hops} &=& - \sum_i \left (\, \epsilon_{_{-Q_i}} \, \sigma^+_i \,
\sigma^-_{i-1} \,+\, \epsilon_{_{Q_i}} \, \sigma^+_{i-1} \, \sigma^-_i \,\right)\,.
\end{eqnarray}
On the other hand, a significant simplification arises for the diagonal terms needed 
for conservation of probability. The former basically count the number of hopping and 
pairing instances at which a given kink configuration can evolve to different ones in a 
single step. Although this would introduce two-body interaction terms of the form 
$\sigma^+_i  \sigma^-_i \sigma^+_{i+1} \sigma^-_{i+1}$ it can be readily verified
that their coefficients will all cancel out so long as the rates involved are those of 
Eq.\,(\ref{rates}), ultimately stemming from detailed balance. Thus, after some brief 
calculations we obtain
\begin{eqnarray}
\nonumber
H_d &=& \sum_i\,\left(\,\epsilon_{_{-P_i}}+\,h_i\,\sigma^+_i \sigma^-_i \,\right)\,,\\
\label{diagonal}
h_i &=& \big( \tanh Q_i + \tanh P_i  + \tanh P_{i+1} - \tanh Q_{i+1} \big)/2\,,
\end{eqnarray}
and no kink interactions, irrespective of disorder in the original couplings. 
Also, it is worth remarking that since ultimately the signs of these latter do not affect 
the  dynamics  [\,Eq.\,(\ref{mapping})\,], at low temperatures and times much smaller 
than $t_B \simeq 2 \min_{\,i} \, \{ \,\left[ 1 - \tanh \left( \, \vert K_i \vert  + \vert 
K_{i-1}\vert\,  \right)  \right]^{-1} \}$ (say with all $J_i \ne 0$), these kink processes 
can then be seen as those of a set  of annihilating random walks in a disordered media 
\cite{Cardy,Odor,us}. Although in this analogy branching walks begin to appear at $t 
\agt t_B$, yet there are some common aspects for both systems at large times (see 
concluding discussion).

To recast the non-diagonal parts of Eqs.\,(\ref{pairs}) and (\ref{hops}) into a  
symmetric representation, we have recourse once more to detailed balance and rotate
each $\sigma^{\pm}_j$ around the $z$ direction using imaginary angles $\phi_j =
i K_j$. So, we consider the diagonal non-unitary similarity transformation $S = \exp\,(\,
{1 \over 2} \sum_i K_i \sigma^z_i \,)$ for which it is straightforward to show that
\begin{eqnarray}
\nonumber
\sigma_i^{\pm}\; \sigma_{i-1}^{\pm} &\to& e^{\pm P_i}\;
\sigma_i^{\pm}\; \sigma_{i-1}^{\pm}\,,\\
\label{symmetrization}
\sigma_i^{\pm}\; \sigma_{i-1}^{\mp} &\to& e^{\pm Q_i}\;
 \sigma_i^{\pm}\; \sigma_{i-1}^{\mp}\,,
\end{eqnarray}
besides keeping unaltered the algebra of $\sigma^+\!,\,\sigma^-\!$. After this 
pseudo spin rotation the above non-diagonal operators therefore transform as
\begin{eqnarray}
\label{s-pairs}
H_{pairs} &\to& -  \frac{1}{2}\, \sum_i\, {\rm sech}\, P_i \, \left (\, \sigma^+_i
\, \sigma^+_{i-1} \,+ {\rm H.\, c.} \,\right)\,,\\
\label{s-hops}
H_{hops} &\to& - \frac{1}{2}\,\sum_i\,  {\rm sech}\, Q_i \, \left (\, \sigma^+_i
\, \sigma^-_{i-1} \,+\, {\rm H.\,c.}\,\right)\,,
\end{eqnarray}
while leaving Eq.\,(\ref{diagonal}) invariant. Thus, we are left with a bilinear 
Hermitian form whose spectrum gap embodies the wanted dynamic exponents
referred to in Sec. I.  In passing, it is worth mentioning that Eqs.\,(\ref{diagonal}), 
(\ref{s-pairs}) and (\ref{s-hops}) also define an anisotropic $XY$ chain under both  
inhomogeneous couplings $J_i \!\!\phantom{.}^{^X_Y} = - \left( {\rm sech} \,Q_i 
\pm {\rm sech} \, P_i \right)$ and transverse fields $h_i$. However, the commutation
algebra of its constituents Pauli matrices,  as well as that of the above raising and
lowering operators, would complicate the analysis. Fortunately, since the parity of
kinks is conserved throughout and all couplings extend just to nearest-neighbors, a
Jordan-Wigner transformation to spinless  fermions $c_i$ \cite{LSM} enables us to
progress further. With the aid of the real symmetric and antisymmetric tridiagonal 
$L \times L$ matrices (with boundaries) $A$ and $B$, defined respectively as

\begin{eqnarray}
\label{AB}
A &=& \left(\begin{array}{cccccc}
h_1 & x_2 & 0 & \cdots  & 0 & -x_1 \\
x_2 & h_2 & x_3 &\; \ddots & \phantom{\ddots} & 0 \\
0    &  x_3 & \phantom{a} \ddots & \ddots & \ddots \;\;\;& \vdots  \\
\vdots & \, \ddots & \ddots & \ddots & x_{L-1} & 0 \\
0 &  \phantom{\ddots}\, &  \ddots \;& \; x_{L-1} & \; h_{L-1} & x_L \\
-x_1\, &\, 0 \,& \cdots &\;\;\;\;0^{\phantom{L^{L^{L^2}}}} & x_L & h_L 
\end{array}\right)\;,\;\;
B =  \left(\begin{array}{cccccc}
0 & y_2 & 0 & \cdots  & 0 & y_1 \\
-y_2 & 0 & y_3 &\;  \ddots & \phantom{\ddots} & 0 \\
0    &  -y_3 &\;\; \ddots & \ddots & \ddots\;\;\; & \vdots  \\
\vdots & \ddots & \ddots & \ddots \;\;& y_{L-1} & 0 \\
0 &  \phantom{\ddots} &  \ddots &  -y_{L-1} & 0 & y_L \\
-y_1\, &\, 0 \,& \cdots &\phantom{--}0^{\phantom{L^{L^{L^2}}}} & -y_L & 0 
\end{array}\right)\,, \\ 
\nonumber
\phantom{void} \\
\label{parameters}
x_i & \equiv & -\frac{1}{2}\;{\rm sech}\, Q_i \;\;,\;\; y_i \equiv 
-\frac{1}{2}\;{\rm sech}\, P_i\,,
\end{eqnarray}
the Jordan-Wigner representation of spins finally lead us to the bilinear fermionic form
\begin{equation}
\label{bilinear}
H =  \sum_i \epsilon_{_{-P_i}} + \,\sum_{i,j} \,\left[\, c_i^\dag \,A_{i,j}\, c_j  \,+\,
\frac{1}{2}\,\left(\, c_i^\dag \,B_{i,j}\, c^\dag_j \, +\,
{\rm H.\,c.}\,  \right)\, \right]\,.
\end{equation}
The boundary matrix elements in $A$ and $B$ are especially aimed for PBC by 
respectively taking account of the contributions of $\sigma^+_L  \sigma^-_1 \to \, 
- c^\dag_L \, c_1$ and  $\sigma^+_L \sigma^+_1 \to - \,c^\dag_L  \, c^\dag_1$. 
Since the number of kinks $\sum_i \!\sigma^+_i \sigma^-_i \equiv \sum_i\! c^\dag_i
\,c_i$ under PBC is always even, only anticyclic conditions must be used in those 
boundary terms.

Following the standard literature \cite{LSM}, this quadratic form can be recasted 
by means of a canonical (unitary) transformation into new  $b_n$-fermions such that 
$\big[\, b_n\, , H \,\big] = E_n\, b_n$, and therefore
\begin{equation}
\label{b-fermions}
H = \frac{L}{2}\,+\, \sum_n\, E_n \Big( \,b^\dag_n \,b_n  - \frac{1}{2} \,\Big)\,.
\end{equation}
Here, the additive constants are obtained by the invariance of the trace of $H$, 
whereas the eigenvalues $E_n$ of its elementary excitations are given by 
the solutions of either of the following two $L \times L$ secular equations \cite{LSM}
\begin{equation}
\label{secular}
\big( A \pm B \big) \,\big(A \mp B\big)\, \varphi^{\pm}_n = 
E^2_n\,\varphi^{\pm}_n\,.
\end{equation}
Because $(A + B)^T = A - B$ the above products are symmetric and hence 
diagonalizable, both being in turn represented by real five-diagonal matrices 
(plus boundaries), as if they were single particle Hamiltonians with hoppings terms 
up to next-nearest neighbors. On the other hand, since by construction $H$ is a 
stochastic operator its vacuum energy always vanishes and therefore all eigenvalues
should be constrained as $\sum_n E_n \equiv L,$ with $E_n > 0$. This is an issue 
which we shall make use of as a consistency test in the numerical diagonalizations of
Sec. III. The wanted relaxation time will then emerge by averaging the spectrum gap 
of this latter secular problem over several of its disorder realizations.

\subsection{Dimerized case}

Before continuing with the numerical analysis of Eq.\,(\ref{secular}) we pause to 
consider briefly the dynamics arising from a periodic array of $J_1$\,-\,$J_2$ couplings
over odd and even bonds, say. Equivalently, the above $P's$ and $Q's$ become 
$P  \equiv  K_1 + K_2\,, \; Q  \equiv  K_2 - K_1$. As already this situation is capable of 
yielding non-universal exponents \cite{Droz}, it is instructive to see how these are 
recovered within the context discussed so far.  We begin by Fourier transforming the 
$c_i$ operators of Eq.\,(\ref{bilinear}) to a set of wave fermions $f_q\, ,\, g_q$ 
\begin{eqnarray}
\nonumber
f_q &=& \sqrt {\frac{2}{L}}\; e^{-\,i\,q/4}\,\sum_{j=1}^{L/2}\, 
e^{-\, i \,q j}\, c_{2j - 1}\,,\\
g_q &=& \sqrt {\frac{2}{L}}\; e^{\,i\,q/4}\,\sum_{j=1}^{L/2} \, 
e^{-\, i \,q j}\, c_{2j}\,,
\end{eqnarray}
with $L/2$ wave numbers 
\begin{equation}
q \in  {\cal Q}^\pm = \left\{ \, \pm 2 \,\frac{\pi}{L}\,,\pm 6 \,\frac{\pi}{L}\,, \cdots\,,
\pm (L - 6) \,\frac{\pi}{L}\,,\pm (L - 2)\,\frac{\pi}{L}\,\right\}\,,
\end{equation}
accounting of the anticyclic conditions referred to above. The $e^{\pm\,i\,q/4}$
phase factors are just aimed at producing a real representation of $H$ when 
expressed in terms of $f$'s and $g$'s.  After introducing the $2 \times 2$ matrices,
\begin{equation}
\nonumber
A_q = \left(\begin{array}{cc} 
h^- \!\!&  \;\; x_{q_{ \phantom {1^L}}} \\
x_q\!\! &  h^+\!\! \end{array}\!\!\!\!\right)\;,\;\;
B_q = \left(\begin{array}{cc} 
0 & y_{q_{ \phantom {1^L}}} \\
y_q & 0^{ \phantom {L}}
\end{array}\!\!\!\!\right)\,,
\end{equation}
parametrized as
\begin{eqnarray}
x_q &=& -\,{\rm sech}\,Q \, \cos \frac{q}{2}\,,\;\;
y_q = -\,{\rm sech}\,P \, \sin \frac{q}{2}\,,\\
\nonumber
h^{\pm} &=& \tanh P \pm \tanh Q\,,
\end{eqnarray}
it is readily found that the  problem splits into subspaces involving only the 
occupation numbers and pairing states of four fermions. Specifically, the dynamics
is decomposed as
\begin{equation}
H = L\, \epsilon_{_{- P}} + \sum_{q \,\in \,{\cal Q}^\pm} H_A (q) 
\, + \, \frac{1}{2} \sum_{q \,\in \,{\cal Q}^+}\! \!\!H_B (q)\,,
\end{equation}
where
\begin{eqnarray}
\nonumber
H_A (q) &=& \big(\, f_q^\dag \,,\, g_q^\dag \,\big)\;  A_q \,
\left(\begin{array}{c} 
f_q \\
g_q \end{array}\right)\,,\\
H_B (q) &=& \big(\, f_q^\dag \,,\, g_q^\dag \,,\,
f_{-q}^\dag \,,\, g_{-q}^\dag\,\big)\; 
\left(\begin{array}{cc} 
0 & B _q  \\
\!\! -B_q & 0
\end{array}\right)\,
\left( \begin{array}{c} 
f^{\dag^{\phantom B}}_q \\
g^{\dag^{\phantom B}}_q   \\
f^{\dag^{\phantom B}}_{-q} \\
g^{\dag^{\phantom B}}_{-q}
\end{array}\!\right)\, + \, {\rm H.\, c.}\;,
\end{eqnarray}
which is the bilinear counterpart of Eq.\,(\ref{bilinear}) in this momentum space. 
Hence, the solution of the secular equation (\ref{secular}) here reduces to the 
diagonalization of either of the products $D_q^{\pm} \, D_q^{\mp}$ of the 
$4 \times 4$ matrices
\begin{equation}
D_q^{\pm} = \left(\!\!\begin{array}{cc} 
A_q & \pm  B_q  \\
\pm  B_{-q} & A_{-q}
\end{array}\right)\,.
\end{equation}
Thus after straightforward algebraic steps and taking the positive roots of the 
associated $E^2_q$ eigenvalues, we are ultimately lead to two elementary 
excitation bands, namely
\begin{eqnarray}
\nonumber
E_q^\pm &=& 1 \,\pm \, \sqrt{\,T^+  \,+\,T^- \cos q\;}\,,\\
\label{bands}
T^{\pm} &=& \left(\,\tanh^2 P \pm \tanh^2 Q\,\right)/2\,.
\end{eqnarray}
Parity conservation of original kinks allows to create only an even number of these 
excitations.  Then, by occupying two $E^-_{q^*}$ levels with $q^* = \pm 2 \pi /L$
when $\vert P \vert > \vert Q \vert$, or alternatively, using $q^* = \pm (L-2) \pi /L$ 
when  $\vert P \vert < \vert Q  \vert$, the spectrum gap  $g = \tau^{-1}$ in the 
thermodynamic limit comes out to be 
\begin{equation}
\label{gap}
g = 2\,\left(\, 1 - \tanh  R \,\right)\,,\;\; 
R = \max \left(\, \vert P \vert\,,\, \vert Q \vert\,\right)\,,
\end{equation}
On the other hand since in equilibrium all kinks become independent, the spin-spin 
correlation length is simply $\xi^{-1} = - {1 \over 2} \ln \left(\,\tanh \vert K_1 \vert \, 
\tanh \vert K_2 \vert\, \right)$ \cite{Droz}, thus growing as $\xi \sim e^{\,2 \vert K_1 
\vert}$ in the low temperature regime, say for $\vert J_2  \vert \ge   \vert J_1 \vert$.
Therefore using Eq.\,(\ref{gap}) within  that latter limit  ($g  \sim  4\, e^{-2 R}$),  
we can now relate these two seemingly independent quantities via the non-universal 
dynamic exponents of Refs.\,\cite{Droz};  that is 
\begin{eqnarray}
\nonumber
\tau & = & \xi^z /\, 4\,,\\
\label{zD}
z &=& 1 \,+ \,\vert J_2/J_1 \vert\;,\; {\rm for}\;\;  \vert J_2 \vert \ge   \vert J_1 \vert\,.
\end{eqnarray}
As expected from the simple but more general arguments given below 
Eq.\,(\ref{mapping}), the role of the coupling signs is irrelevant.

{\it Scaling regimes}.--
Finally, and in preparation for the finite-size scaling comparisons of Sec. III under both
discrete and continuous disorder, it is helpful to consider also the spectrum gap of 
finite chains. Expanding the lower band of Eq.\,(\ref{bands}) to second order around
either of the $q^*$ minima referred to above, and considering afterwards the scaling
regime  $L \to \infty,\, \xi \to \infty$ while holding $x \equiv \xi /L$ finite, it can be 
easily verified that the gap decay exhibits the crossover
\begin{equation}
\label{crossover}
g \propto \cases{ \xi^{-z}\, \;\;\;\;\;\;\;\;\, {\rm for}\, \;\; x \ll 1/\pi\,, \cr
\xi^{\,2-z} / L^2\,\;\;\;\;\;\; {\rm otherwise}\,.}
\end{equation}
Evidently, this implies a typical Arrhenius dependence \cite{Privman} for the 
relaxation time at low $T$'s, that is  $\tau \propto \exp \,( 2\,b /T )$ for $T / \vert J_1 
\vert \ll 1$, though now with energy barriers $b$ fixed by the interplay between
temperatures and sizes, namely
\begin{equation}
\label{barriers}
b = \cases{ \vert J_2 \vert + \vert J_1 \vert\;\;\;
{\rm for}\, \;\;  T / \vert J_1 \vert \gg  1/ \ln \sqrt L  \,, \cr\cr
\vert J_2 \vert - \vert J_1 \vert \;\;\; {\rm otherwise}\,.}
\end{equation}
Moreover, both finite-size and finite correlation length corrections to these 
crossovers can also be estimated by keeping next leading order terms in 
Eq.\,(\ref{bands}). So, these finite effects are readily found to scale as
\begin{equation}
\label{scaling}
g \,\xi^z  =  4 +  4 \pi^2 \left( 1 - \frac{\,\pi^2}{3\,L^2}  \right)
x^2 \,+\,{\cal O} \left( 1 / \xi^z \right)\,.
\end{equation}
We will get back to these issues later on in the context of Figs.\,\ref{binary} and 
\ref{FSS} of Sec. III.

\section{Numerical results}

Turning to the diagonalization of Eq.\,(\ref{secular}), in what follows we consider the
cases of binary and Gaussian probability distributions of disorder realizations, both 
taken site-independent. The coupling concentration ($p$) of the former is such that
$P (J) = (1-p)\, \delta_{J, J_1} + p \,\delta_ {J,J_2}$, whereas the latter is parametrized
as usual by an average $\langle J \rangle$ and variance $\sigma^2 \equiv \langle J^2 
\rangle - \langle J \rangle^2$.  As the temperature is varied we focus on samples 
with correlation lengths close to their averaged disorder values, so as to minimize
the dispersion of relaxation times resulting from finite-size effects. In the discrete 
case this is just equivalent to using a constant number of coupling types, i.e. the 
correlation length is fixed as $\xi_p = - \left[\, (1-p)\, \ln \left( \tanh \vert K_1 \vert 
\right) +p\, \ln \left( \tanh \vert K_2 \vert \right) \,\right]^{-1}$. For the continuous
case, in turn we select and diagonalize only those samples having $\xi^{-1} = 
- \sum_i \ln \left( \,\tanh \vert K_i \vert \,\right) / L \simeq - \ln \left \langle \, \tanh 
\vert K \vert\, \right \rangle$, which amounts to work with disorder fixed (or nearly
fixed) end-to-end spin-spin correlations.
%
\begin{figure}[htbp]
\centering
\includegraphics[width=0.38\textwidth]{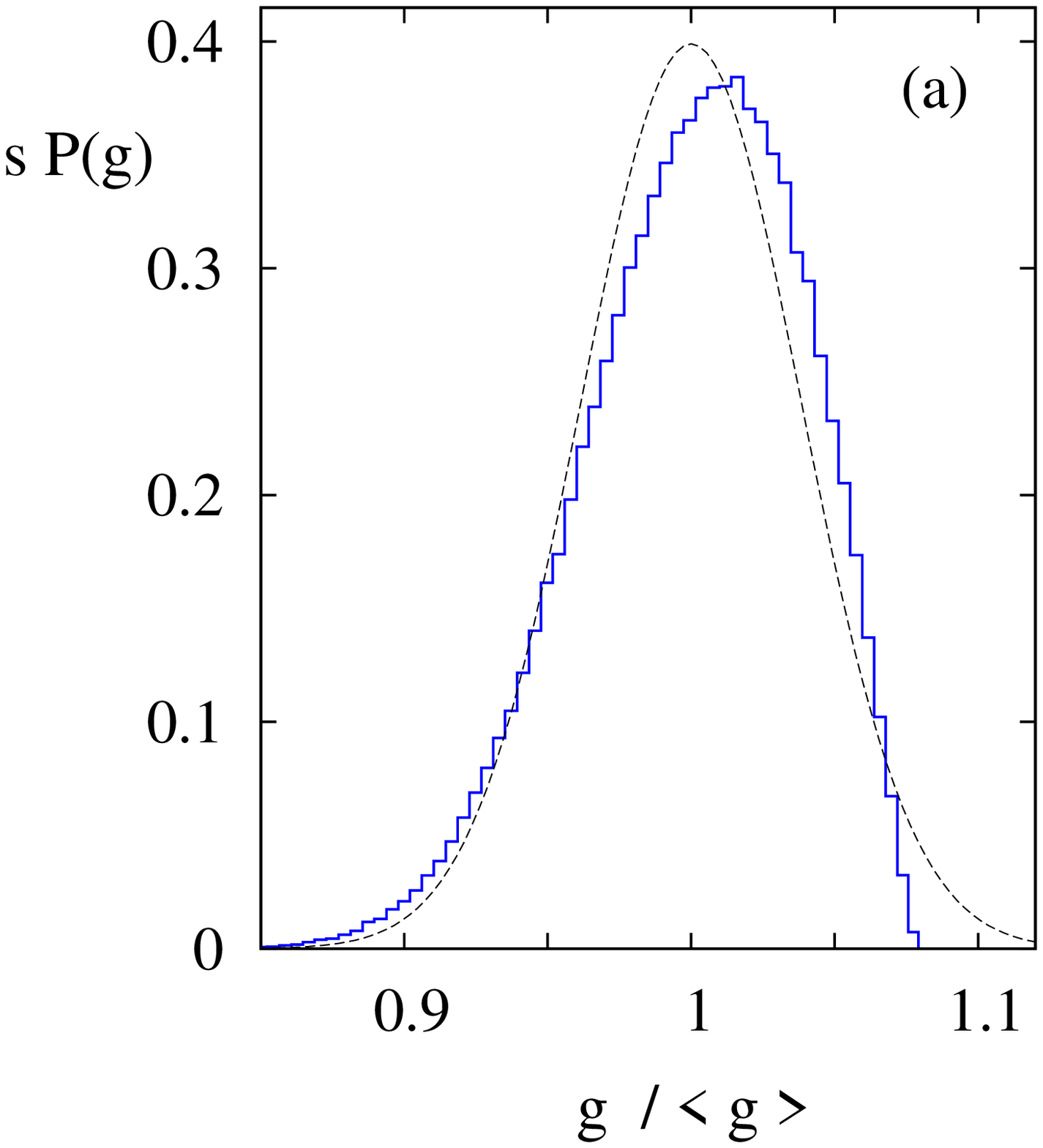}
\includegraphics[width=0.38\textwidth]{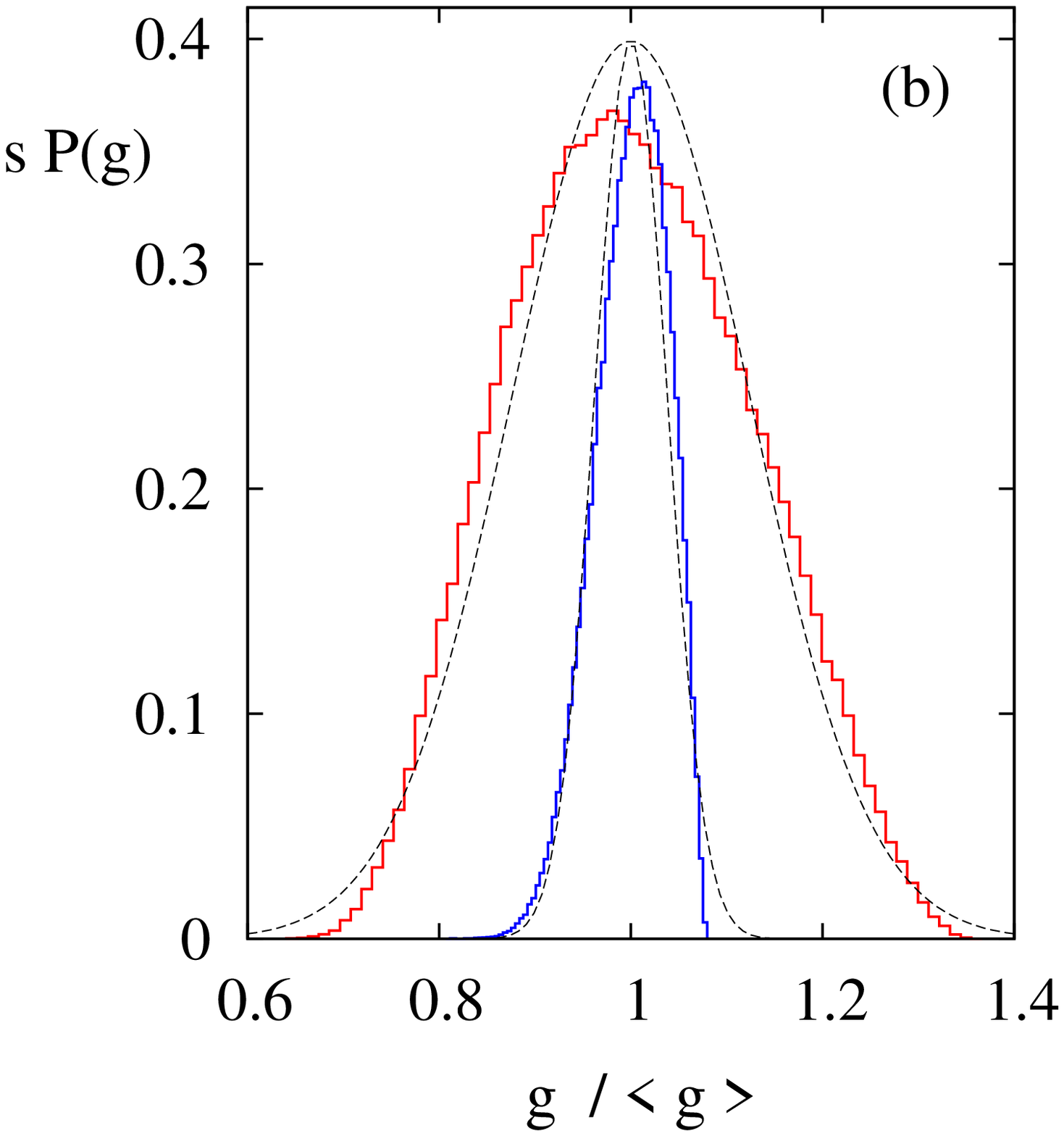}
\vskip 0.5cm
\caption{(Color online) Gap distributions over $3 \times  10^5$ samples of $500$ 
spins after binning data in 70 intervals using (a) binary and, (b) Gaussian disorder 
realizations with fixed correlation length ($\xi \simeq L/2$). For comparison, dashed
lines display Gaussian fittings with same mean and standard deviation ($s$) of 
otherwise skewed distributions. (a) illustrates  the typical  narrow dispersion $(s /
\langle g \rangle \simeq 0.04)$ observed in the binary case (here,  $T/\vert J_1 \vert =
0.36,\, \vert J_2 / J_1 \vert = 2$, and $p = 0.5$). In (b) an almost identical quantitative 
behavior is displayed for $\langle J \rangle = 2$ with $T/ \langle J \rangle = 0.3$ and
$\sigma / \langle J \rangle = 0.15$ (innermost curve), whereas by contrast the case 
$T/ \langle J \rangle = 0.2$, $\sigma / \langle J  \rangle = 0.3 > \bar\sigma_c$ (see 
text)  but similar $\xi$, develops a more symmetric and broader dispersion $(s / 
\langle g \rangle \simeq 0.12)$.}
\label{distributions}
\end{figure}
%

In Fig.\,\ref{distributions} we show the typical probability distributions of gaps 
arising from this simple procedure. Already for $500$ spins it is seen that standard 
gap deviations turn out to be fairly small. In particular, the abrupt decrease of these
distributions above $g \agt \langle g \rangle$ possibly signals the presence of natural
upper bounds. However, for Gaussian disorders with $\sigma/\langle J \rangle \agt 
0.22$ (see below) but holding $\xi$ approximately constant (that is to say, choosing
a temperature so as to keep comparable late domain sizes), this negative skewness 
is smeared out and the distributions widen significantly. As we shall see, at the level
of average gaps this latter issue will be reflected in terms of a rather different decay
of $\langle g \rangle$ with the correlation length. In averaging these gaps we used
chains of up to 3000 spins for which the sampling had to be substantially reduced
(down to 20 realizations). This is due to, in part, an $L^3$ increase of the 
diagonalization time of standard routines \cite{NR}, as well as to a slowing down
of convergence within low temperature or large $\xi$ regimes. The smallness of the
spectrum gap for these regions and chain lengths also precluded us from using
Lanczos-based algorithms \cite{Lanczos}, as their convergence became even slower.
Ultimately, these issues set the practical limit to the idea of singling out arbitrarily
large relaxation times, as intended to in Sec. I. 

\subsection{Average binary gaps}

\begin{figure}[htbp]
\vskip -2.1cm
\hskip -1.1cm
\includegraphics[width=0.575\textwidth]{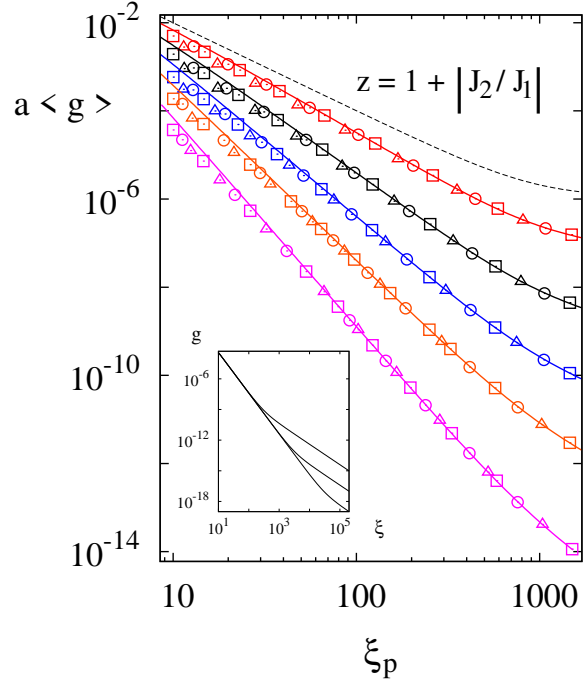}
\vskip -2.3cm
\caption{(Color online) Average gaps over 20 realizations of 3000 couplings drawn 
from binary distributions $(1-p)\, \delta_{J, J_1} + p \,\delta_ {J,J_2}$. Squares, 
circles, and triangles refer respectively to concentrations $p = 1/4,\, 1/2,$ and 3/4 
with correlation lengths $\xi_p$ as referred to in the text. Solid lines stand from top 
to bottom for dimerized cases with $\left\vert  J_2 / J_1 \right\vert = 1.5,\, 2,\, 2.5,\, 3,$
and 3.75 ($\xi \equiv  \xi_{p = 1/2}$). Within the scaling region $\xi \alt  L/\pi$, these
closely follow the non-universal power law decays of their disordered counterparts,
the gaps of which, for displaying purposes, have been normalized by different 
amplitudes.  The inset illustrates the finite-size crossover mentioned in 
Eq.\,(\ref{crossover}) for the dimerized situation using $L = 10^3\!, 10^4\!,10^5$ 
(top to bottom), and $\left\vert J_2 / J_1 \right\vert = 3$ (also, see Fig.\,\ref{FSS}).
For comparison, the uppermost dashed line in the main panel denotes the uniform
case $z = 2$.} 
\label{binary}
\end{figure}

\begin{figure}[htbp]
\vskip -2.3cm
\hskip -0.1cm
\includegraphics[width=0.575\textwidth]{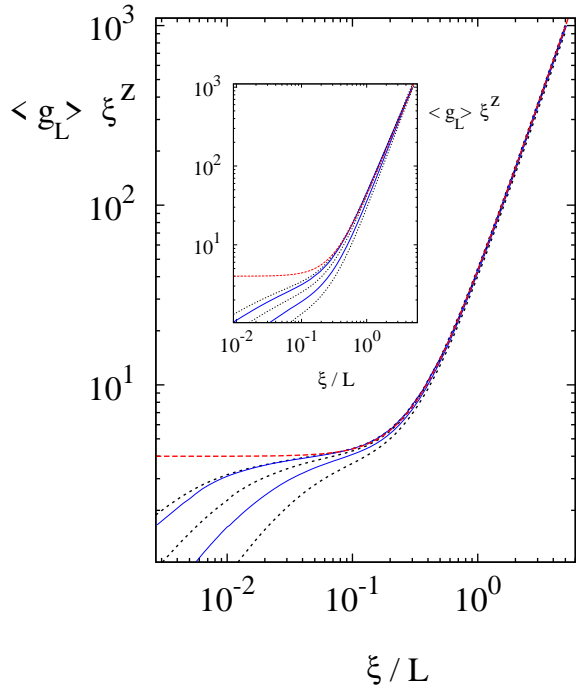}
\vskip -2.4cm
\caption{(Color online) Finite-size scaling regimes of average gaps and correlation
lengths for a binary coupling disorder with $p = 0.5$ and $\left\vert J_2 / J_1 \right
\vert = 2.5$ (main panel). From left to right doted and solid lines stand in turn for 
$L = 3000,\, 2000,\,1000,\,500,$ and 200. The data collapse in the scaling region 
$\xi / L \agt 0.3$ is attained upon setting $z = 1 + \left\vert J_2 / J_1 \right\vert$. 
For much smaller values of  $\xi / L$ finite-size departures show up with the same 
sign as those of Eq.\,(\ref{scaling}), the thermodynamic limit of which is denoted 
by the dashed line. Similarly, the inset displays the scaling regimes of a Gaussian 
distribution with $\sigma /\langle J \rangle = 0.15$ upon choosing $z \simeq 2.61$. 
For comparison with the above thermodynamic limit, all finite size data were rescaled
by a common factor.}
\label{FSS}
\end{figure}

Nonetheless, the relative errors of $\langle g \rangle$ yet remain small and clear 
trends emerge as $\xi$ is varied. Fig.\,\ref{binary} illustrates the resulting average
decays for several parameter values of the binary distribution, all with $\vert  J_2 \vert
> \vert J_1 \vert$. It can be observed that within the very same scaling regime $\xi \alt
L/\pi$ referred to in Eq.\,(\ref{crossover}) for the plain dimerized chain, our results 
suggest a similar gap decay i.e. $\langle g \rangle \propto 1/\xi^{\,1 \,+ \,\vert J_2/J_1 
\vert}$, and irrespective of disorder concentrations \cite{Droz2, Robin}.
%
%
Departures of data from this regime are equally followed by the exact solution of the 
alternating-bond chain, the  finite-size aspects of which are depicted in the inset. 
Furthermore, as is shown in Fig.\,\ref{FSS}, above $\xi/L \sim 1/\pi$ the numerical 
diagonalizations also reproduce the universal scaling function contained in the
thermodynamic limit of Eq.\,(\ref{scaling}) (appart from an overall $p$ dependent
amplitude), which therefore evidence the same crossover decay, i.e. $\langle g \rangle
\propto \xi^{2-z} / L^2$, for the binary distribution.
%
%
Alike the dimerized case, deviations from scaling are negligible within this region (all
sizes yielding slopes $\simeq 2$), though as $\xi$ decreases subdominant $1/\xi^z$
corrections bring about breakdowns. Just as in Eq.\,(\ref{scaling}), these happen to
be of the same sign and more severe at the smaller sizes. 

\subsection{Average Gaussian gaps}

\begin{figure}[htbp]
\vskip -2.7cm
\hskip -1.1cm
\includegraphics[width=0.575\textwidth]{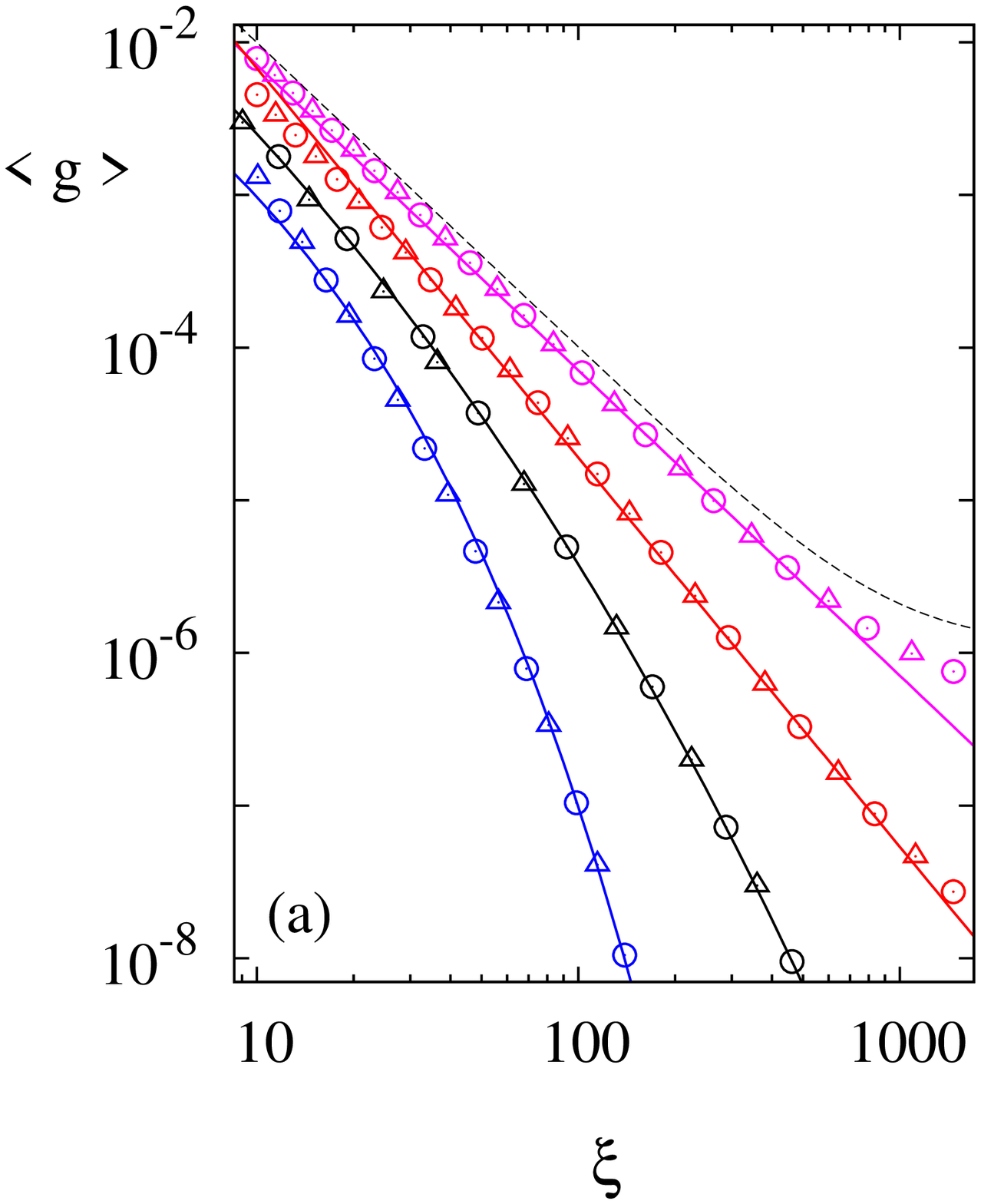}
\hskip -1.69cm
\includegraphics[width=0.575\textwidth]{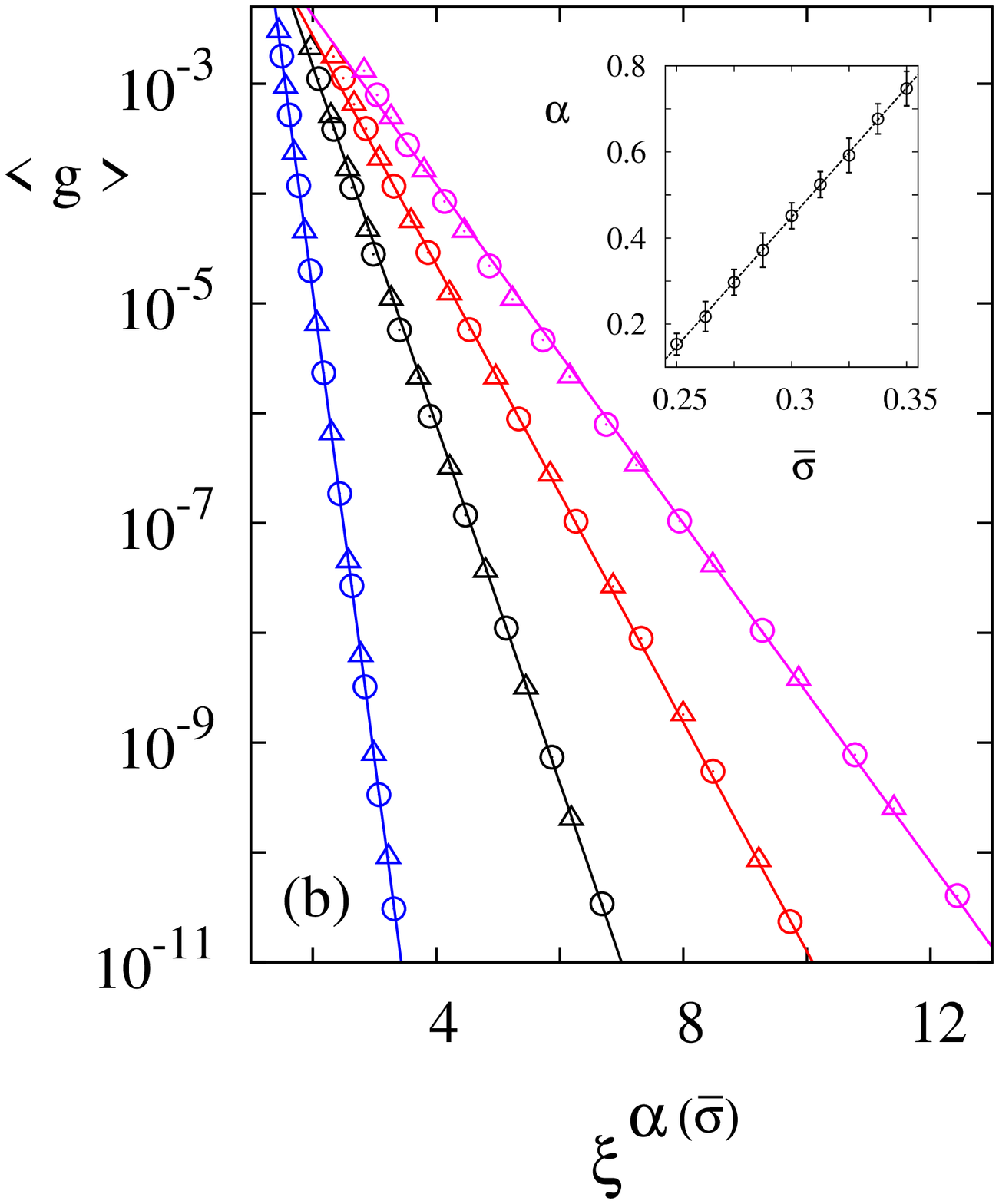}
\vskip -2cm
\caption{(Color online) Average spectrum gap for a Gaussian distribution of couplings
over 20 samples of 3000 spins, all realization having in turn a common correlation
length. In the main panels circles and triangles stand respectively for $\langle J
\rangle = 1$ and 2. From left to right each data set referrs to (a) $\bar \sigma \equiv
\sigma/ \langle J \rangle = 0.3, \,0.25, \,0.2, \,0.1$ and, (b) $\bar \sigma = 0.25, \, 
0.275, \, 0.2875, \, 0.3$. For $\bar \sigma < \bar \sigma_c \simeq 0.22$, in (a) data 
are fitted with non-universal power laws, whereas above such critical value all data 
closely follow stretched exponential decays [\,Eq.\,(\ref{stretched})\,] with stretching
exponents $\alpha$ estimated as in the inset of (b). For comparison, dashed line in (a)
denotes the ordered case $\sigma = 0$.}
\label{Gaussian}
\end{figure}

Similar scaling features appear also for Gaussian disorder provided $\sigma$ is 
held small enough (see below), although finite-size departures become even more 
prominent for $\xi \alt L$, as is observed in inset of Fig.\,\ref{FSS}. In evaluating 
average gaps for this latter type of disorder, and in parallel with the behavior of the 
gap distributions referred to in Fig.\,\ref{distributions}b, two decay regimes are now
obtained according to whether the relative standard deviation $\bar\sigma \equiv
\sigma/ \langle J \rangle$ is smaller or greater than a certain threshold $\bar
\sigma_c$. This is displayed in Fig.\,\ref{Gaussian}a where we show some trends
for 3000 spins using several variances for $\langle J \rangle = 1$ and 2. As long as 
$\bar\sigma <  \bar\sigma_c$ all gaps exhibit usual (but non-universal) power law
decays $\propto 1/ \xi^z$ with $z$ increasing slightly between 2 and $\sim$2.65 
as $\bar\sigma$ increases. The incipient crossovers observed at $\xi \agt 1000$ 
correspond to the scaling regimes already alluded to in the inset of Fig.\,\ref{FSS},
so there $\langle g \rangle \propto \xi^{2-z}$. 
%
%

By contrast, however, for $\bar\sigma > \bar\sigma_c$ a much faster decay emerges
over the whole range of accessible correlation lengths, whereas the scaling relation
put forward in Fig.\,\ref{FSS} no longer holds. We direct the reader's attention to 
Fig.\,\ref{Gaussian}b where it turns out that the non-linear least-squares fitting of 
the stretched exponential form 
\begin{equation}
\label{stretched}
\langle g \rangle \sim A\, \exp\,\left( -\,a\,\xi^{\alpha} \right)\,,
\end{equation}
is able to follow very closely all numerical diagonalizations. After estimating the 
stretching exponents of several variances it was found that above a threshold of 
$\bar \sigma_c \sim 0.22$ up to $\bar \sigma \sim 0.35$, they increase with $\bar 
\sigma$ as $\alpha \sim 6.01\,\left(\,\bar \sigma - \bar\sigma_c\right)$ as is shown in
the inset of Fig.\,\ref{Gaussian}b. In nearing that critical variance however, it is 
difficult to distinguish the power law decay obtained before from this new behavior, 
so the crossover from one $\sigma$-regime to the other turns out to be smooth. 

Let us mention that it would be important to elucidate whether this 
conjectured non-algebraic decay actually extends beyond the regions displayed
in Fig.\,\ref{Gaussian}b, e.g. $\xi \alt$ 1500, 270 for $\bar\sigma = $ 0.25  and
0.3 respectively. However, above those correlation scales the numerical gaps get 
progressively smaller [\,in turn resulting from the even smaller squared gaps of the
secular problem (\ref{secular})\,], which brings about a much slower and erratic
convergence of diagonalization routines \cite{NR}. 

\section{Concluding discussion}

Recapitulating, we have constructed a fermionic bilinear representation of the 1D 
Glauber dynamics with nearest-neighbor random interactions. This corresponds to 
the symmetric version [\,Eq.\,(\ref{symmetrization})\,] of the kink pairing and 
hopping processes schematized in Fig.\,\ref{kinks}. The relaxation time of these 
latter was reduced to the evaluation of the single-particle spectral gaps of the secular 
problem given in Eq.\,(\ref{secular}), the diagonalization of which ultimately led us
to non-universal forms of domain growth. Some of these were expected and others 
not foreseen. 

The case of binary disorder lent itself more readily for comparisons with the soluble 
alternating chain of same  coupling types (Sec. II\,A). Although the former situation
has no exact solution, based on decimation procedures as well as on simple diffusion
arguments found in the literature \cite{Droz2}, the numerical coincidence of dynamic
exponents with those of the soluble case should come as no surprise 
(Fig.\,\ref{binary}). In addition, the  scaling function of Eq.\,(\ref{scaling}) reproduced 
our binary data just above regions displaying incipient departures from the actual 
exponents (Fig.\,\ref{FSS}, and rightmost part of Fig.\,\ref{binary}). Because of this 
scaling robustness, the same barrier values involved in the crossover of Arrhenius 
times [\,Eq.\,(\ref{barriers})\,] might be expected to also hold in finite chains with 
binary disorder regardless of their bond concentrations. In passing, it is worth pointing
out  that such manifestation of finite-size effects has been observed experimentally in
 single-chain magnets \cite{Coulon}.

Some of these scaling aspects seem to also apply for Gaussian disorder realizations
with small relative variances (inset of Fig.\,\ref{FSS}), for which non-universal but yet 
power-law forms of gap decay were obtained (uppermost data in Fig.\,\ref{Gaussian}a).
However, for $\sigma/ \langle J \rangle \agt 0.22$ a crossover to new regimes with 
much faster and non-algebraic decays showed up, whilst at the more detailed level
of gap distributions this brought about a significant increase of dispersion
(Fig.\,\ref{distributions}b). The stretched exponential form put forward to fit our 
data (Fig.\,\ref{Gaussian}b) lent further support to the former hypothesis which, by
inverting Eq.\,(\ref{stretched}) and recalling the large scale equivalences referred to
in Sec. I \cite{Privman, Menyhard}, is tantamount to a dynamics of domain scales
growing asymptotically as $\sim (\ln t)^{1/\alpha}$. Curiously, this rather slow form
of phase-ordering kinetics is reminiscent to that conjectured and studied in late 
coarsening stages of higher dimensions with bond disorder \cite{Huse, Lippiello}.
In turn,  this behavior may be also viewed as that corresponding to a diverging 
dynamic exponent, somewhat analogous  to activated scaling \cite{Fisher} in random 
transverse-field magnets. The  reason for such divergence is similar to that found in 
disordered reaction-diffusion processes thought of as annihilating random walks in a 
Brownian potential \cite{Cardy,Odor,us}. The idea is that the equations of motion 
implicit in $\left[\, H\,,\, c_l\,\right] = \sum_m \left( \,A_{l,m} \,c_m\,+ \,B_{l,m} \, 
c^\dagger_m \right)$ [\,see Eqs.\,(\ref{AB})-(\ref{bilinear})\,], can be mapped into a 
biased random walk of phase steps \cite{us} whose first-passage probability to 
traverse a  distance  of order $\ln \tau$ sets a relationship between the dynamic 
exponent and the nonuniversal -disorder dependent- ratio $D / b$ (diffusion constant 
and bias of the random walk of phases). The limit of zero bias would then be related to 
an unbounded  $z$-exponent and to a crossing over to the stretched exponential 
regime conjectured  above \cite{us}. As mentioned earlier on, at large times the 
analogy with annihilating random walks does not strictly apply but the reasoning in 
terms of steps of random phases goes along similar lines \cite{us} (though here the 
mapping of the equations of motion couples the phase steps of two fields).

Alongside these growth law evaluations, it would be interesting to also address the
issue of scaling and superuniversality (or the lack thereof), in two-time quantities
such as autocorrelation and autoresponse functions \cite{Lippiello, Henkel}. That
research line might well be further investigated with the fermion approach discussed
in this work. Contrariwise, the effect of external fields \cite{Forgacs}, whether random
or not, would be bound to adopt rather cumbersome expressions had it been written
in terms of the kink representation. For analogous reasons, the treatment of Glauber 
dynamics in higher dimensions remains beyond the scope of this approach.

\section*{Acknowledgments}

M.D.G  acknowledges support of CONICET and ANPCyT, Argentina, under Grants 
PIP 1691 and PICT 1426.



\end{document}